# Understanding and Fixing Complex Faults in Embedded Cyberphysical Systems

## PREPRINT

*Alexander Weiss\*, Smitha Gautham\*\*, Athira Varma Jayakumar\*\*, Carl Elks\*\*, D. Richard Kuhn\*\*\*, Raghu N. Kacker\*\*\*, Thomas B. Preusser\**

\* Accemic Technologies
\*\* Virginia Commonwealth University
\*\*\* National Institute of Standards and Technology

Understanding fault types can lead to novel approaches to debugging and runtime verification. Dealing with complex faults, particularly in the challenging area of embedded systems, craves for more powerful tools, which are now becoming available to engineers.

## MISTAKES, ERRORS, DEFECTS, BUGS, FAULTS AND ANOMALIES

Embedded systems are everywhere, and they present unique challenges in verification and testing. The real-time nature of many embedded systems produces complex failure modes that are especially hard to detect and prevent. To prevent system failures, we need to understand the nature and common types of software anomalies. We also need to think about how mistakes lead to observable anomalies and how these can be differentiated according to their reproducibility. Another key need for efficient fault detection is the comprehensive observability of a system. This kind of understanding leads to the principle of "scientific debugging".

In everyday language, we use a number of words such as bug, fault, error, etc. inconsistently and confusingly to describe the malfunctioning of a software-based system. This also happens to the authors of this paper in their normal life unless they pay strict attention to their choice of words. Therefore, we would like to start with a brief clarification based on the terminology used by the "IEEE Standard Classification for Software Anomalies" [1].

- If coders notice a mistake themselves, it is called an *error* ("a human action that produces an incorrect result "[1]).
- If the tester is the first to notice an anomaly ("a product does not meet its requirements" [1], it is called a *defect*. After confirmation by the developer, it becomes a *bug*.

- If an end user finds the problem ("manifestation of an error" [1]), we have a *fault*.

Figure1 illustrates the semantics of anomalies. The term "anomaly" may be used to refer to errors, defects, bugs as well as faults. It refers to something that deviates from what is expected or normal with respect to the required behavior, ideally defined by a system specification.

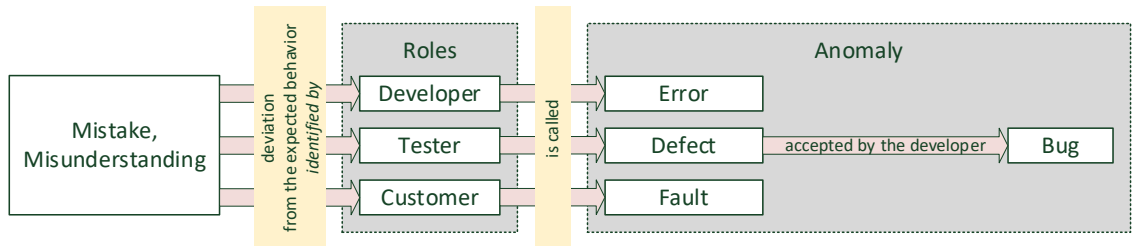

Figure 1: Semantics of Mistakes, Errors, Defects, Bugs, Faults, Anomalies etc.

# REPRODUCIBILITY OF ANOMALIES

For the engineer's ability to eliminate a bug or a fault (= debugging), its reproducibility is crucial. This property is, therefore, an essential classification criterion for anomalies. A deterministic manifestation is the repeatable occurrence of an anomaly under a well-defined, but possibly not yet understood, set of conditions. Such a manifestation is also called a Bohrbug, named after Bohr's deterministic atom model. To be consistent with the terminology established in related work, we will use the terms "Bohrbug", "Mandelbug" etc. instead of the more consistent but odd-sounding wording "Bohr anomaly", "Mandel anomaly" etc.

If the underlying causes for an anomaly are so complex and obscure that it appears to be non-deterministic, we speak of a Mandelbug (named after the chaotic Mandelbrot set). Race conditions seen in concurrent programs are common examples of Mandelbugs [2].

The literature defines subclasses of Mandelbugs:

- Aging-related bugs occur in long-running systems due to error conditions caused by the accumulation of problems such as memory leakage, propagating rounding errors or unreleased files and locks. A typical example for an aging-related bug is the software fault in the Patriot missile-defense system (see text block).
- Anomalies that seem to disappear or alter their behavior when looked into are called Heisenbugs, named after the uncertainty principle described by the physicist Werner Heisenberg, which is often informally conflated with the probe effect.

It might seem that the predominant class of faults should be complex Mandelbugs. This is not the case: In practice, there is a surprisingly high proportion of Bohrbugs. An example is given by Grottke et al. [3] who analyzed the software faults for 18 JPL/NASA space missions. Out of 520 software faults, 61.4% were clearly identified as Bohrbugs, and 36.5% as Mandelbugs (4.4% of which were aging-related).

> **PATRIOT'S FATAL FLAW**
>
> *On February 25, 1991, during the first Gulf War, a Patriot air defense system near Dhahran, Saudi Arabia, failed in its task of detecting and intercepting an Iraqi scud missile. The result was that it hit American barracks unhindered, killing 28 soldiers and injuring about 100.*
>
> *The reason was an "aging-related bug" in the weapon control computer. An inaccuracy caused by rounding a floating-point variable continued to propagate. With increasing operating time, the target detection became so inaccurate that the trajectory of attacking missiles was wrongly judged as harmless* [4].

Nonetheless, Mandelbugs increasingly gain importance as more complex systems are being developed, and seemingly non-deterministic fault patterns occur more frequently with concurrency and parallelism in multicore systems. For example, Ratliff et al. [5] showed that Mandelbugs tend to involve more interacting factors than the deterministic Bohrbugs, with roughly one additional, indirect factor on average.

## THE ANATOMY OF AN ANOMALY

The possible effects of a mistake are illustrated in Figure 2. Our example system traverses through a sequence of states, $z_1 \ldots z_6$, which are characterized by the internal variables, $i_1 \ldots i_4$. The system produces the observable outputs, $e_1$ and $e_2$. Each program execution step computes an update to the internal variables and the outputs to produce.

If the executed program contains a mistake, the resulting state might not be as expected. In this case, we speak of an activated mistake and a resulting infected state where an anomaly has manifested.

By the transition from state $z_1$ to $z_2$, two code segments with mistakes are executed. One of them (code A) computes a wrong state of the internal variable $i_3$, the other (code B) produces a wrong observable output $e_2$. The latter is a textbook manifestation of a Bohrbug as long as the anomaly can be reproduced under a well-defined, but possibly not yet understood, set of conditions.

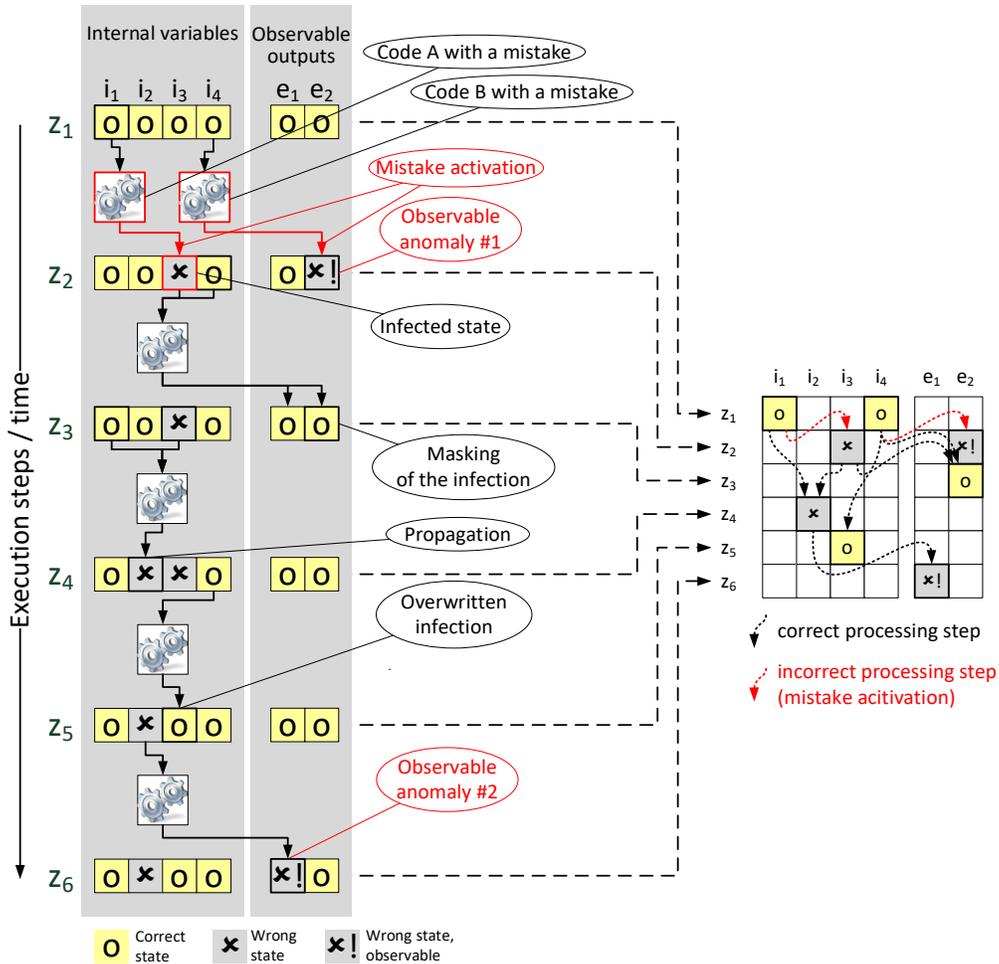

Figure 2: A defective program execution as a succession of states (inspired by [4]). Left-hand-side: extended depiction of states and transitions. Right-hand-side: compacted depiction of the same scenario (inspired by [5])

A typical Mandelbug scenario (as caused by code A) is an anomaly that only changes an internal variable. This may not be detectable easily. Worse, the actual root infection can potentially be overwritten and masked regularly. The path of propagating and overwriting infections can cause many headaches. In our example, during the transition to state $z_4$, the wrong internal variable $i_3$ causes a wrong assignment of $i_2$. When $i_3$ is overwritten in $z_5$, the track record of this originally wrong variable is lost before the error is exposed in $z_6$. By this time, no indication remains to point to the software defect in switching from $z_1$ to $z_2$.

If the transitions are processed in a multicore system, there is an increased chance for a more chaotic manifestation of the anomalies. To avoid this nightmare scenario, comprehensive monitoring capabilities are essential.

## THE DEBUGGING PROCESS

The process of understanding the underlying cause of an anomaly, i.e. the identification of the mistake, and fixing the problem is called debugging. Starting from an observed anomaly, a hypothesis (i.e. a testing theory) that narrows the space of possibilities for its cause is developed. The next step is to develop an experiment to test this hypothesis. If the hypothesis is supported, either the detected mistake can be understood and fixed or the hypothesis can be further refined. If the hypothesis was false, a new hypothesis has to be developed.

It is obvious that observability is a crucial factor for an efficient debugging process. The technology and tools to cope with complex faults in embedded systems are now becoming available to engineers.

## THE OBSERVATION TOOLBOX

### Printf() Debugging (Figure 3a)

Named after the `printf()` function from the C programming language, is the most basic form of observation in the debugging process. Source code is manually instrumented to output debug information. Unfortunately, this approach can have a massive impact on timing behavior, or even introduce unintended synchronization in concurrent programs when the same output console is shared. This creates a perfect setup for running into otherwise unrelated Heisenbugs. Additionally, if the information of interest changes, the software must be adapted and recompiled. Even though this approach is archaic, it is still in use. In the worst case, its tediousness and time-consuming nature may threaten project goals if no other more advanced debug method is available.

### Start/Stop Debugging (Figure 3b)

This common approach is based on the direct control over a program's execution. Halting execution at selected breakpoints, the reached program state may be inspected and analyzed in detail. However, this approach has major drawbacks for analyzing complex transient anomalies:

1. Directly controlling the execution progress of a program changes its timing behavior. This makes it hard to reproduce anomalies whose manifestation depends on timing. In Cyber Physical Systems (CPS), where the software is controlling physical actuators, halting the control could even cause physical damage.
2. It typically only allows stepping forward. Breakpoints have to be chosen thoughtfully to be early enough to reflect the root cause of an observed anomaly. This typically leads to running the software over and over again in an effort to trace back the original manifestation of an anomaly.
3. Due to the cyclic debugging fashion, the system behavior is required to be fundamentally deterministic to enable the observation and investigation of an anomaly. This is hard or impossible to achieve in parallel or in real-time programs with dynamic asynchronous input data.

### Omniscient Debugging (Figure 3c)

These debuggers are also known as back-in-time or reversible debuggers. They record the whole or parts of a program's execution. This enables going back in time by reconstructing the execution history and the corresponding program contexts. This approach most naturally aligns with tracing back the root cause of an anomaly starting from an unexpected observation.

The challenge posed by omniscient debugging is in capturing the execution trace. Software instrumentation is an option. Due to its behavioral feedback into the monitored application, engineers however, face a delicate dilemma having to trade off tracing detail against the faithful behavioral representation of the program. Compromises performing more coarse-grain tracing as on the level of function calls are common. Again, within critical control environments or with hardware in the loop, even such observation compromises are often simply not tolerable.

The key technology to enable truly non-intrusive and yet detailed omniscient debugging is **Embedded Trace**. Dedicated on-chip modules directly integrated into the monitored CPU capture, encode and emit the execution trace. Arm® CoreSight™ [6] or Intel® Processor Trace [7] are prominent implementations of this technology.

Embedded Trace implementations rely on an aggressive sophisticated trace data compression. A naïve encoding just of all the addresses of the instructions executed by a 1 GHz CPU coming from a 32-bit address space would produce 32 Gbps of data. Economically feasible implementations reduce this bandwidth demand to 1 Gbps and below. This is achieved by only encoding diversions from the default sequential control flow and by pruning all information that can be inferred from the executed application binary. Other, often optional, information channels may add to the actual bandwidth demand. Providing timing information or data trace [8] is particularly challenging in this respect.

Omniscient debugging in practice is severely challenged by trace data volume and limited by storage capacity. Two capture paths are common:

- System memory: This option competes over memory bandwidth and space within the monitored system, so it impacts the system's performance, and timing behavior and is limited to short observation time spans only.
- External capture: An external tracing device captures trace data from a designated interface for later offline processing. While this avoids possibly disruptive feedback into the monitored system, implementations are limited by the buffer capacity. In practice, trace clips of up to a few seconds are possible.

As both of these state-of-the-art approaches rely on buffering, they impose similar analytic limitations on the backend trace interpretation. Firstly, the escalation of a root cause into an observed anomaly must fit into the trace window to avoid a multi-run reconstruction. Also, the anomaly must be reproducible reliably and predictably to allow its capture by a quick trace snapshot. The only assistance natively offered by the various embedded trace architectures is a few primitive triggers that allow filtering the trace at its source. They are too simple to capture complex interrelated conditions, and too few to allow for a defensive guarding against multiple hypothetical escalation paths.

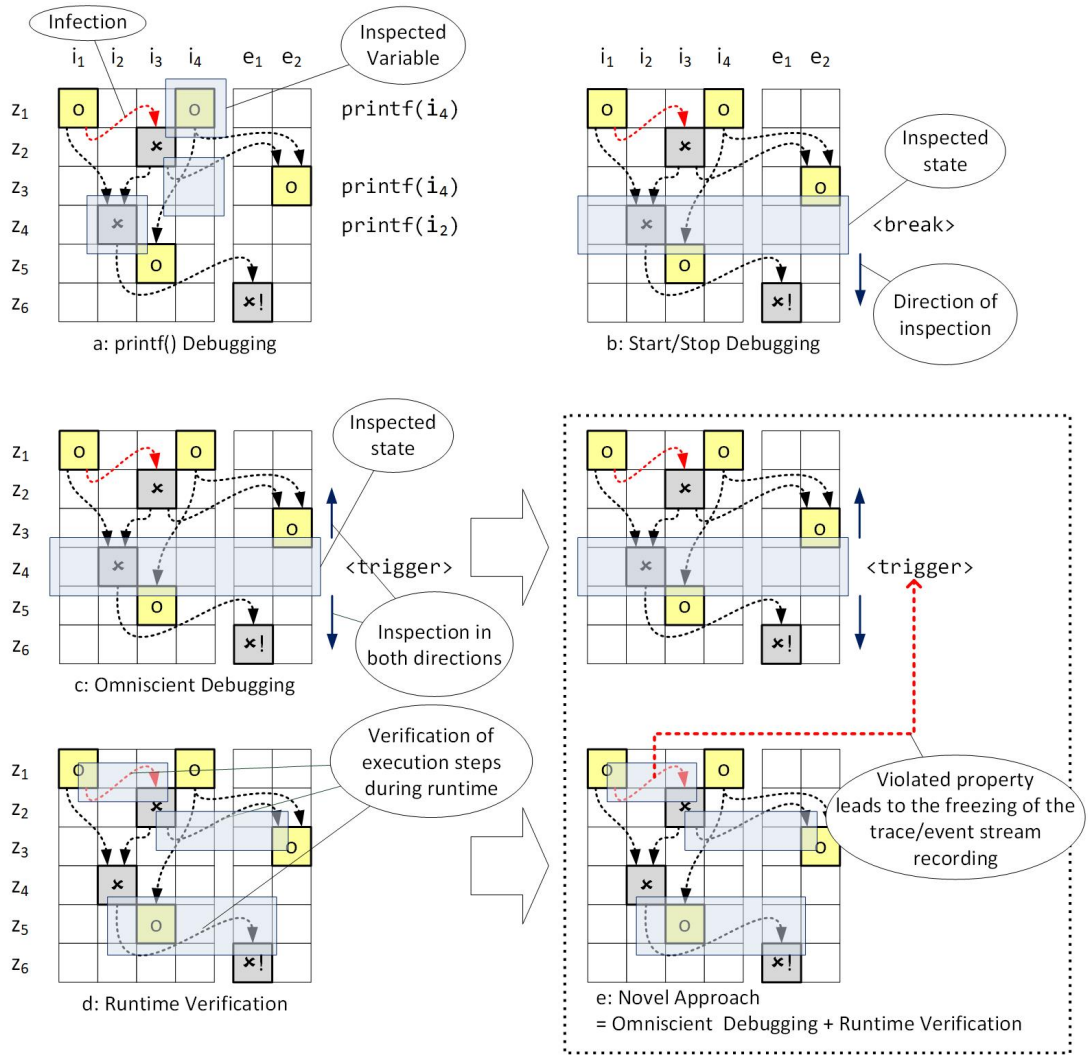

**Figure 3: (a) printf() Debugging, (b) Start/Stop Debugging, (c) Omniscient Debugging, (d) Runtime Verification (e) Novel approach combing the best of the Omniscient Debugger and the Runtime Verification approaches.
This illustration follows the representation introduced in Figure 2.**

## Runtime Verification (Figure 3d)

One promising technique that enables immediate detection of infected states is Runtime Verification (RV) [9], [10] a dynamic testing technique that is often called *light-weight formal verification*. It sits in between traditional dynamic testing methods discussed in previous sections and formal mathematical proof techniques. Unlike a proof, it is not complete or exhaustive but it provides more guarantees of correctness than debug style testing. In general, runtime verification makes use of a *monitor* that observes the execution behavior of a target system.

To verify correctness, a monitor uses a *specification of acceptable behavior,* often derived from natural language requirements and translated into temporal logic formulas to allow reasoning over time-sensitive sequences and states. These formulas reside in the monitor. Execution trace information (i.e. states, function variables, decision predicates, etc.) is extracted directly from the target system and forwarded to the monitor where the temporal logic is elaborated with the trace data for an on-the-fly validation of system behavior as shown in Figure 3d. As such, runtime verification can detect violations of predefined properties immediately. Runtime verification extends debugging and testing by ensuring that important system properties are continuously checked during operational phases. Runtime verification can be used to detect Bohrbugs, as well as Mandelbugs that may be masked as seen in the scenario in Figure 2. Detecting Mandelbugs by runtime verification is especially beneficial as these bugs can elude development phase testing.

That said, a key concern for runtime verification is the observability of the system operation at time intervals of interest. This is often attained by software instrumentation with its limitations described previously.

Recently, new approaches for minimally intrusive runtime verification have produced significant capabilities to verify and test very complex program behavior as discussed below.

## RUNTIME VERIFICATION FOR TESTING

We often see testing complemented with runtime verification as the two provide an excellent verification means to detect complex faults (Falzon & Pace [11], Colombo[12]). Especially, model-based designs provide a good framework to exploit synergies between testing and runtime verification. Additionally, runtime monitors can act as test oracles, ensuring that critical properties hold true during testing. Runtime verification is supported by testing frameworks such as MathWorks Simulink and ModelJUnit, a model-based testing framework written in Java [12].

Runtime verification languages need more expressiveness to specify complex properties that emerge from multi-threaded interactions in complex embedded systems. Signal Temporal Logic (STL), Event Calculus, and Metric Temporal Logic (MTL) [13] are some of the RV languages able to express specifications for monitoring complex embedded systems and CPSs. Another recent approach is Stream-based Runtime Verification (SRV), which combines event processing and runtime verification. An example of a runtime verification tool that performs stream-based runtime verification is TeSSLa [14]. TeSSLa is designed for specifying properties where timing and sequencing is critical. It supports timestamped events and a declarative programming style, which is well suited for expressing specifications. Furthermore, runtime verification tools such as TeSSLa and Copilot allow the automatic synthesis of executable code monitors often realized in C or HDL (Hardware Description Language) code, directly from the runtime verification language [13] [14].

As we stated in the Observation Toolbox section, most microprocessors today provide embedded trace in some form to support either start-stop debugging or omniscient debugging. These embedded trace features (sometimes called On Chip Debugging cores) can significantly reduce the intrusiveness of software instrumentation to support runtime verification. For example, the Arm® CoreSight™ architecture [6] features an Instrumentation Trace Macrocell (ITM) that can be used to output custom trace messages with minimal intrusion. Compared to traditional printf() debugging, which takes several milliseconds of CPU time, ITM takes very few clock cycles (100's of nanoseconds) to emit custom trace data. However, these embedded trace cores are not straightforward to interface to or program for software instrumentation purposes. Therefore, it remains a challenge for certain applications.

## Novel Approaches (Figure 3e)

The approaches and methods discussed above are established in practice but are largely separate from one another. To make substantial gains in testing efficiency and reliability, we need novel methods that are both non-intrusive and more integrative.

The great practical challenge of omniscient debugging is the limited size of trace buffer memory combined with the general problem of identifying the manifestation of anomalies within the trace data stream. The trigger logic built into a processor's execution trace infrastructure is too simple to validate meaningful higher-level behavioral execution models. It rather takes designated runtime verification to deliver precise and complex triggering.

The key enabler of such a capable runtime verification is the instantaneous full behavioral disclosure of the system operation. This demands that captured execution trace data is not stored away but rather decompressed and interpreted on-the-fly. Its decoded information can be leveraged to keep a digital twin in sync, which models the relevant behavior of the observed system. Violations of behavioral constraints are then reliable triggers for freezing the recording of raw execution trace data or refined higher-level events in a ring buffer. The captured trace clip describes precisely the pathway taken by the system towards this violation ("save on trigger", Figure 3 (e)).

Decompressing and interpreting the processor execution trace online and at the rate of its emission is technically challenging. Field-Programmable Gate Arrays (FPGAs), which are user-programmable integrated circuits that enable the implementation of custom circuitry on off-the-shelf chips, are well suited for such applications. For example, the CEDARtools® platform [15] utilizes large FPGAs allowing tremendous parallelism to manage such decompression and interpretation of traces online. Spatially designated processing modules decode the trace data stream into messages, resolve them against a model of the executed binary, extract a stream of relevant higher-level events from the reconstructed control flow, and subject this event stream to a constraint's validation. The latter is achieved by an array of dataflow processors that are programmed in the specification language TeSSLa using temporal logic. Violations of the specified constraints freeze the relevant, most recent trace and event history for in-depth analysis by the engineer.

## USE CASE: SMART SENSOR FOR NUCLEAR POWER APPLICATIONS

The authors recently conducted a study on applying software systematic testing process on an embedded smart sensor device, representative of the type found in Nuclear Power generation applications [16].

Nuclear power generation facilities worldwide are steadily trending towards "aging infrastructure" with the average age of a Light Water Reactor in the United States at about 37 years. The need to modernize these generation facilities with software based I&C systems and smart sensors is high priority for the industry. These systems have to be tested and validated to very high levels of safety assurance (IEC 61508-SIL 4) to ensure as reasonably possible that no software faults exist.

The VCU smart sensor is a real time embedded device used in a Cyber Physical Systems context – monitoring physical states of a nuclear reactor. The smart sensor hosts a real-time lightweight operating system that executes several concurrent threads associated with sensing barometric pressure, temperature and maintaining device health ([17], Appendix B). The systematic testing methodology we employed is called "Pseudo-Exhaustive" testing which is built around t-way combinatorial testing, partitioning, boundary value analysis, and MC/DC path analysis [18]. We applied the systematic testing of the software following the well-known unit test, integration test, and system test paradigm.

A Testbed architecture was built using state of the art test automation tools necessary to conduct the real-time systematic testing experiments on the smart sensor [17]. The three major tools used in our testbed are (1) Razorcat's TESSY, (2) National Institute of Standards and Technology (NIST) ACTS tool, and (3) Keil Interactive Debugger. TESSY, developed by Razorcat is an automated testing tool for safety-critical embedded systems software [19].

Among the findings from this study, which reinforce the observation toolbox discussed earlier, is that analyzing test failures to find an underlying cause can be a significant effort depending on the nature of the bug. Detecting failures at the interface level is a preferable starting point, but to find causality and fix a bug requires circumspection of the internal software behaviors [9]. Analyzing test failures depends on a number of factors, but in our experience three factors are significant; observability, controllability and complexity - with respect to target system software execution. Observability is the ability to witness software state and interaction conditions at the temporal resolution needed to perceive the anomalous behavior. As software becomes more complex and multi-threaded, demands on the observability of executing code become more critical.

Complexity is not a goal of software engineering; it's a consequence of the way processor technology has evolved to become more computationally powerful. Many of the challenges we encountered testing the smart sensor embedded software were directly related to the complexity of software interactions among software threads. This complexity is evident for anyone who has developed or tried to read labeled control flow graphs and data flow graphs of multi-threaded software artifacts. The number of control and data flow paths decides the number of tests required to get complete test execution coverage of software units.

Figure 4a shows a simple example of a labeled function call graph from the smart sensor software with the edge labels defined. Each node in this directed graph denotes unique functions within the software and the edges within the graph are unidirectional in nature indicating a function at the *edge head* being called by another function at the *edge tail*. As seen in this call graph example, there are direct/unconditional function calls and function calls made conditionally based on single or multiple conditions. The edge between the functions 'mcu_init' and 'i2cStart' is labeled with conditions 'X1, X2' denoting that the function 'mcu_init' calls the function 'i2cStart' when either 'ARIES_USE_I2C1' or 'ARIES_USE_I2C2' macros are defined, i.e. when any of the two I2C channels usage is enabled in the software. The call graph also indicates that the function 'mcu_init' calls the functions 'gptStart', 'chHeapInit' and 'gptStartContinuous' functions unconditionally as the edges between these functions are labeled 'D', which stands for direct call.

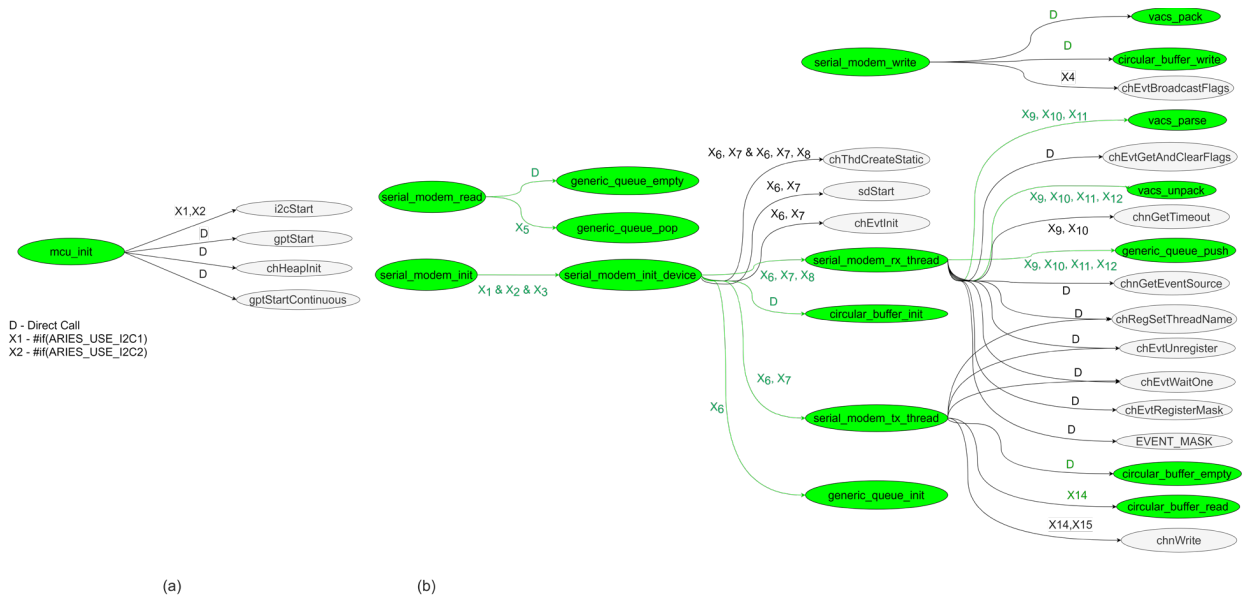

**Figure 4: (a) Simple Function Call Graph of mcu_init, (b) Complex Function Call graph of Serial_modem thread**

As an example of the complexity that can arise with simple multi-threaded applications, Figure 4b is a complete call graph of one of the several threads 'serial_modem' within the multi-threaded smart sensor software. The comprehension, analysis and testing of software with complex call graphs becomes extremely challenging as the data flow and control flow couplings are difficult to track – even with the aid of advanced debuggers, like IDA Pro [20].

Real-time embedded systems need to be verified both in the temporal domain and value domain. Testing these function call graph execution sequences for all valid data flows ended up being more complex and challenging than combinatorial testing as it involved understanding and handling the periodicity and order of invoking tasks by the Real-Time Operating System scheduler. In order to accurately emulate the function call order as per the call graphs, software testers needed to develop a detailed knowledge of the function call and control flow sequencing. Debugging test failures identified during sequence testing is very challenging, as it involves stepping across multiple functions, watching internal variables, outputs and arguments passed across multiple functions, and verifying the order of external function calls.

Our study revealed a number of bugs that had remained latent in the code for years. We found that combinatorial t-way testing helped detect and identify a few Bohrbugs quickly (only two-way interaction was needed). These Bohrbugs were easily reproducible and repeatable with multiple tests resulting in 'Infinity' and 'Not A Number' outcomes.

A harder to detect and identify bug was a Heisenbug scenario that appears rarely in the situation of a hardware fault or cyber-attacks leading to an invalid serial bus (e.g. I2C serial bus) input. This bug was caught during combinatorial testing with corner case inputs. This buffer overflow which is a Heisenbug, is caused due to two colluding factors - usage of a longer I2C input buffer size and missing bounds check on received data. Such software anomalies that occur due to the combination of multiple root causes in the software are more difficult to track down with traditional testing and debugging methods.

Lastly, a Mandelbug that was uncovered during integration testing was most challenging to detect and analyze. In this case, a deadlock situation arose when two software components that interacted with each other were not allowed to proceed as they were both waiting for inputs from each other. These kinds of anomalies are seen in multi-threaded software with complex state machine interactions such that the threads/functions progress their states based on inputs from other components (both hardware and software). The root cause of these kinds of deadlock behaviors can be very difficult to observe, detect and analyze as it requires carefully tracing back the entire history of state progressions within both the software components under the given input conditions.

This use case was a modest software-based device – a smart sensor, and it presented testing challenges for the testers. We reasonably conclude that with the proliferation of advanced high-performance embedded systems (e.g. multi-core processors, and heterogeneous processors) there is a strong need for powerful approaches and supporting tools to manage complexity of embedded software testing.

## CONCLUSIONS

Embedded systems are becoming much more common in daily life, and better ways of finding and preventing failures are essential. The complexity posed by Cyber-Physical Systems present grand challenges to testing and verification. The state of practice for embedded software is at a point where new methods and novel techniques are needed to adequately test these critical systems. Advancements in understanding the nature of complex faults, and applying this understanding in maturing testing and verification, make it possible to build embedded Cyber Physical Systems that are safe and secure.

*Acknowledgment - Parts of this research were supported by the US Department of Energy, Idaho National Laboratory through the Light Water Reactor Sustainability program under contract FP 217984. Other Parts of this research received funding from the European Union's Horizon 2020 research and innovation program under grant agreement No. 732016 and Federal Ministry of Education and Research of Germany in the KMU Innovative project CoCOSI (project number 01IS19044). **Disclaimer:** Any mention of commercial products in this paper is for information only; it does not imply recommendation or endorsement by NIST.*

## ABOUT THE AUTHORS

ALEXANDER WEISS is the co-founder and chief executive officer of Accemic Technologies, Kiefersfelden, Germany. His research interests include cyber physical systems runtime analysis, software verification, functional safety, and cyberattack detection. Weiss received his Dr.-Ing. in computer science from TU Dresden. Contact him at aweiss@accemic.com.

SMITHA GAUTHAM is a Ph.D. student in the Department of Electrical and Computer Engineering, Virginia Commonwealth University, Richmond, Virginia, USA. Her research interests include the design and assessment of heterogeneous runtime verification architectures, runtime safety and security monitors for cyber physical systems, and model-based design assurance and verification. Gautham received her M.S. in electrical engineering from Virginia Commonwealth University. Contact her at gauthamsm@vcu.edu.

ATHIRA VARMA JAYAKUMAR is a research assistant in the Department of Electrical and Computer Engineering, Virginia Commonwealth University, Richmond, Virginia, USA. Her research interests include the dependability and security of cyber physical systems, fault and cyberattack injection, model-based design assurance and verification, and the validation of embedded software. Jayakumar received her M.S. in computer engineering from Virginia Commonwealth University. Contact her at jayakumarav@vcu.edu.

CARL R. ELKS is an associate professor in the Department of Electrical and Computer Engineering, Virginia Commonwealth University, Richmond, Virginia, USA. His research interests include the analysis, design, and assessment of dependable embedded cyber physical systems of the type found in critical infrastructure. Elks received his Ph.D. in electrical engineering from the University of Virginia in 2005. He is a Member of IEEE. Contact him at crelks@vcu.edu.

D. RICHARD KUHN is a computer scientist in the National Institute of Standards and Technology Computer Security Division, Gaithersburg, Maryland, USA. His research interests include combinatorial methods for software verification and applications to autonomous systems as well as empirical studies of software failure. Kuhn received his M.S. degree in computer science from the University of Maryland, College Park. He is a Fellow of IEEE. Contact him at kuhn@nist.gov.

RAGHU N. KACKER is a mathematical statistician in the National Institute of Standards and Technology Applied and Computational Mathematics Division, Gaithersburg, Maryland, USA. His research interests include combinatorial methods for software testing, artificial intelligence testing, and autonomous systems. Kacker received his Ph.D. degree in statistics from Iowa State University. Contact him at raghu.kacker@nist.gov.

THOMAS B. PREUSSER leads the research team at Accemic Technologies, Dresden, Germany. His research interests include massively parallel compute acceleration and data processing in heterogeneous FPGA-based systems as well as computer arithmetic. Preusser received his Ph.D. (Dr.-Ing.) in computer engineering from TU Dresden. Contact him at tpreusser@accemic.com.